\documentclass[preprints,article,accept,moreauthors]{Definitions/mdpi} 
\firstpage{1} 
\pubvolume{1}
\issuenum{1}
\articlenumber{0}
\pubyear{2026}
\copyrightyear{2026}
\datereceived{ } 
\daterevised{ } % Comment out if no revised date
\dateaccepted{ } 
\datepublished{ } 
\Title{A Six-Quadrupole Diamagnetic Suspension Architecture for Ultra-Sensitive Space Accelerometry}

\Author{Marco Pisani $^{1}$\orcidA{}, Edoardo Dalla Ricca $^{1}$, Carlo Paolo Sasso $^{1}$\orcidB{} and Massimo Zucco$^{1}$  }%$^{1,}$*}

\AuthorNames{Firstname Lastname, Firstname Lastname and Firstname Lastname}

\address{%
$^{1}$ \quad Istituto Nazionale di Ricerca Metrologica}

\corres{Correspondence: m.pisani@inrim.it; Tel.: +39-011-3919-966}

\abstract{This paper proposes a novel inertial sensor 
based on a solid diamagnetic cubic test mass passively suspended by a symmetric configuration of six magnetic quadrupoles. Designed specifically for microgravity environments, this system eliminates the mechanical noise and hysteresis inherent in elastic suspensions, as well as the complexity associated with electrostatic suspension systems. We provide an analysis of the magnetic intrinsic noise floor imposed by eddy current damping, alongside ground-based preliminary experimental results demonstrating potential sensitivities in the order of the $10^{-10} \text{ m/s}^2/\sqrt{\text{Hz}}$ regime.}


\begin{document}

\section{Introduction}

Acceleration measurements play a crucial role in a wide range of space applications, ranging from fundamental physics experiments, to satellite geodesy and accurate orbit determination and correction for navigation satellites \cite{hines2022optomechanical,weber2022application,visser2013thermospheric,touboul1999electrostatic}.
In a microgravity environment, spaceborne accelerometers operate by measuring the relative displacement of a nominally free-floating test mass with respect to its spacecraft housing. Non-inertial accelerations acting on the spacecraft, such as those arising from solar radiation pressure or residual atmospheric drag, are imparted to the spacecraft, while the test mass remains in near-free fall. The resulting differential motion therefore produces a measurable displacement between the test mass and its surrounding housing.
To realize this measurement in practice, the test mass must be physically or virtually constrained to remain within a defined working volume. Consequently, the core architecture of any space accelerometer is defined by two essential elements: the constraint (suspension) system and the displacement readout system.\\
The simplest form of suspension is a mechanical elastic element (spring), characterized by a specific stiffness constant $k$. Depending on the suspension geometry, the test mass may remain free to move along between one and six degrees of freedom (DoFs). For example, a blade spring constrains all but one translational DoF, resulting in a single-axis accelerometer (such as ISA aboard BepiColombo \cite{IAFOLLA2010}), whereas a coil spring can provide compliance in all six DoFs, each characterized by a different effective stiffness. %This spring–mass principle is also employed in micro-electromechanical systems (MEMS), which provide a compact and cost-effective alternative to conventional accelerometers. In these devices, a miniature proof mass is suspended by compliant microfabricated springs, demonstrating acceleration sensitivities on the order of $10^{-2} \mathrm{m/s^2/\sqrt{Hz}}$ \cite{middlemiss2016measurement,prasad2018portable,li2016novel}.\\
Regardless of the implementation, the mechanical sensitivity of a spring–mass accelerometer, defined as the displacement produced per unit applied acceleration, is fundamentally determined by the restoring stiffness and the proof mass. When operated well below the mechanical resonance, the proof-mass displacement is  given by:

\begin{equation}
    x=a/\omega_0^2
    \label{eq:acc}
\end{equation}
where $a$ is the applied acceleration and $\omega_0$ is the natural angular frequency of the test mass. Therefore, for a given mass, maximizing sensitivity requires minimizing the spring constant $k$.\\
To achieve this ultra-low stiffness, state-of-the-art space accelerometers predominantly utilize electrostatic suspensions \cite{christophe2015new,bai2017research}. Through active control loops, the electrostatic forces between the test mass and fixed electrodes mounted on its surrounding housing are controlled to create a virtual spring, allowing the stiffness to be reduced to extremely low values. While highly effective, the main disadvantage of electrostatic suspensions is the strict requirement for a continuous, closed-loop active control system, which relies on the knowledge of the test mass position within its housing.\\
Meeting this requirement places stringent demands on the readout system. Capacitive sensing is the most widely adopted solution, in which the position of the test mass is inferred from changes in the capacitance between the test mass and the surrounding electrodes \cite{josselin1999capacitive}. Alternatively, optical interferometric readouts have demonstrated even higher displacement sensitivities \cite{LISA-INTERFEROMETER,pisani2018accelerometer}.\\
In this paper, we introduce a novel paradigm for test mass constraint systems: a fully passive, contactless architecture based on the magnetic suspension of a diamagnetic test mass within a system of magnetic quadrupoles. By exploiting the large diamagnetic susceptibility of solid bismuth and the high degree of symmetry provided by the magnetic housing, we demonstrate the feasibility of generating a stable three-dimensional magnetic potential well. This approach eliminates mechanical dissipation and noise associated with conventional spring-based suspensions, and removes the complexity of active feedback schemes typically required in electrostatic sensing systems. The resulting architecture provides a simple solution for the development of next-generation ultra-sensitive space accelerometers.

The use of diamagnetism for the levitation of small masses and for the development of tilt and acceleration sensors has been previously demonstrated, but primarily in the presence of gravity, which provides a natural mechanical constraint, and for very small test masses. Moreover, none of these approaches has employed a symmetric three-degree-of-freedom (3-DoF) structure \cite{geim1999magnet,chen2026triaxial,xu2026design,homans2025experimental,wang2025ultra}.\\

%%%%%%%%%%%%%%%%%%%%%%%%%%%%%%%%%%%%%%%%%%
\section{Working Principle, Architecture and Gap-Tuning}

The core of the sensor consists of a cubic bismuth test mass enclosed within a three-dimensional magnetic trap. The test mass is suspended inside a housing formed by a point (central) symmetric configuration of six magnetic quadrupoles, providing contactless confinement along all three spatial axes. 
\subsection{Magnetic Configuration}
\begin{figure}
            \centering
            \includegraphics[width=1\linewidth]{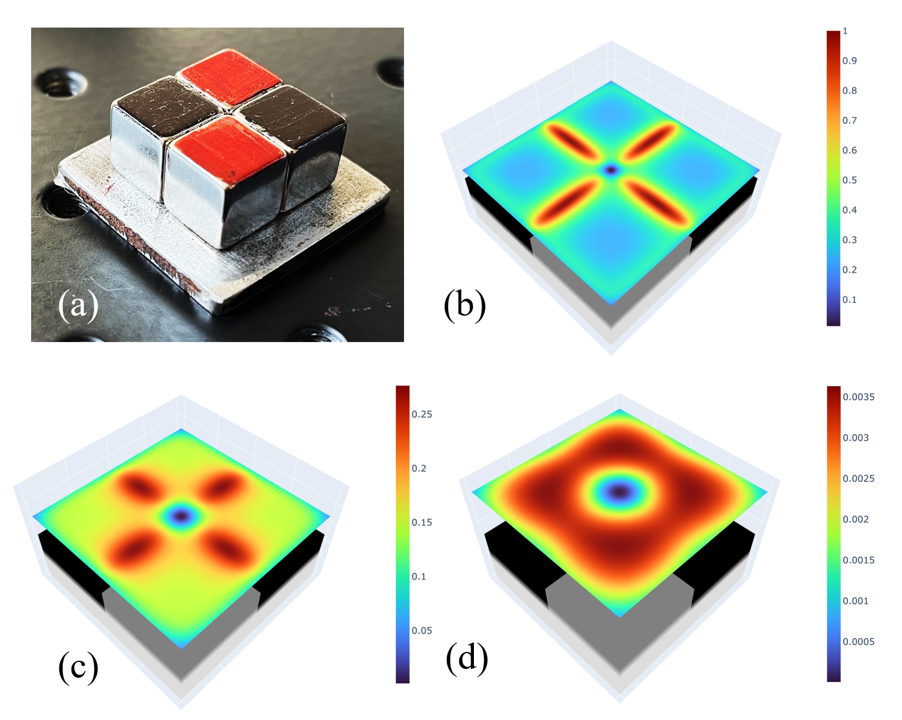}
            \caption{(a): Picture of the single quadrupole element. The poles are colored for clarity; (b, c and d): graphical representation of the value of $B   \nabla B$ for different distances from the quadrupole surface: respectively 1 mm, 2 mm and 4 mm. The false color scale is normalized to the maximum value of the 1 mm case}
            \label{fig:fieldgradient}
\end{figure}

 As a baseline for the following analysis we have considered a cubic test mass having \textit{l} = 25 mm and \textit{m} = 153 g. Each quadrupole is composed of four cubic NdFeB permanent magnets (10 mm side length) arranged in a checkerboard magnetization pattern (N–S/S–N), as shown in Figure \ref{fig:fieldgradient}(a). To enhance the inward concentration of the magnetic flux, the rear face of each quadrupole is backed by a 30×30×2 mm soft iron plate which has also the purpose of shielding the magnetic flux outwards. This configuration generates a theoretical zero-field point at the geometric center of the trap, surrounded by a region of extremely large magnetic field gradient ($\nabla B$). Since the diamagnetic force is proportional to the product of the magnetic field and its spatial gradient ($F \propto B\nabla B$), the bismuth test mass experiences a repulsive force from each quadrupole (see Figure \ref{fig:fieldgradient} (b, c and d)). The arrangement of six quadruples will act as a multidimensional virtual spring, or 3-dimensional potential well,  pushing the mass towards the center of the housing.
 \subsection{Gap tuning}   
 The gap is defined as the clearance between the surface of the  test mass and the magnetic poles. We have taken as a baseline a nominal gap of 2 mm, achieving a compromise between restoring force and allowable displacement. Because the magnetic field generated by the quadrupoles decays rapidly with distance, increasing the gap results in a significant reduction in both the restoring force and the effective magnetic spring stiffness. An advantage of this configuration over conventional mechanical springs is the ability to tune the system stiffness by varying the gap between the levitated mass and the surrounding structure. Reducing the stiffness correspondingly lowers the resonance frequency, potentially bringing it into the mHz regime thus enhancing the accelerometer’s sensitivity in a controllable manner. Moreover, as discussed in the following section, increasing the gap also lowers the instrument’s intrinsic thermal-magnetic noise floor.

\section{Fundamental Noise Analysis and bismuth anisotropy}

%The movement of the conductive bismuth mass through the intense magnetic gradient induces eddy currents within its volume. These currents dissipate energy via Joule heating, introducing a magnetic damping coefficient $\Gamma_{mag}$. This dissipation is inextricably linked to a stochastic dissipative force noise given by:

%\begin{equation}
%    F_{damping} = -\Gamma_{mag}v
%\end{equation}

%where $v$ is the test mass velocity.\\
The ultimate resolution limit of the accelerometer is governed by the Fluctuation-Dissipation Theorem (FDT) \cite{callen1951irreversibility} of electromagnetic nature, given that mechanical linkages are not present. The movement of the conductive bismuth mass through the intense magnetic gradient induces eddy currents within its volume. These currents dissipate energy via Joule heating, introducing a magnetic damping coefficient $\Gamma_{mag}$. This dissipation is inextricably linked to a stochastic dissipative force noise, with an amplitude spectral density of acceleration noise given by:

\begin{equation}
    S_{a,th}^{1/2} = \sqrt{\frac{4k_BT\Gamma_{mag}}{m^2}}
\end{equation}

where $k_B$ is the Boltzmann constant, $T$ the thermodynamic temperature and  $m$ is the mass of the test mass.\\
The power dissipated by eddy currents scales with the square of the magnetic-field spatial gradient, consequently, the associated thermal acceleration noise scales linearly with the magnetic-field gradient. The acceleration noise is obtained by first evaluating the viscous damping due to eddy currents induced in the bismuth cube as it moves through the magnetic-field gradients, which reaches approximately 
75 T$/$m  for a 2 mm gap and 40 T$/$m for a 4 mm gap. 
The resulting total damping coefficients are 
$\Gamma_{mag} \approx 0.133 \text{\ N s$/$m}$ and $\Gamma_{mag} \approx 0.037 \text{\ N s$/$m}$, respectively. Applying the Fluctuation–Dissipation Theorem at $T =300$ K, these damping coefficients yield thermal acceleration-noise spectral densities of:

\begin{equation}
    S_{a,th}^{1/2} \approx 3.07 \times10^{-10} \mathrm{m/s^2/\sqrt{Hz}}
\end{equation}
for the 2 mm gap, and

\begin{equation}
    S_{a,th}^{1/2} \approx 1.61 \times10^{-10} \mathrm{m/s^2/\sqrt{Hz}}
\end{equation}

for the 4 mm gap.
Increasing the gap from 2 mm to 4 mm approximately halves the magnetic field gradient, reducing the thermal magnetic noise floor by a factor of two.\\
 Beyond this, the strong anisotropy of the diamagnetic susceptibility of the bismuth crystal is a critical property to be addressed 
 \cite{shoenberg1936magnetic}. If the test mass consists of large single crystals, this anisotropy results in asymmetric restoring forces within the otherwise symmetric magnetic housing. To ensure an isotropic macroscopic magnetic response, the formation of large crystals must therefore be suppressed. Manufacturing techniques such as rapid solidification or mechanical agitation during casting promote the formation of a fine-grained polycrystalline microstructure with randomly oriented crystallites, effectively averaging out the intrinsic anisotropy.\\
On the other hand, the intrinsic thermal-magnetic noise can be reduced by replacing the solid crystalline test mass with a sectorized structure (such as smaller cubes separated by dielectric layers), to reduce eddy currents. An alternative and potentially optimal approach is to employ powder metallurgy, in which the bismuth is sintered from electrically isolated grains. This approach offers the possibility of addressing both limitations simultaneously: a test mass fabricated from pressed, randomly oriented bismuth particles with insulating coatings would provide near-isotropic magnetic properties while substantially increasing the macroscopic electrical resistivity thus reducing the eddy current damping at the material level.

\section{Experimental Setup and Preliminary Results}

\begin{figure}[!h]
 		\centering
        \subfloat[]{
 		\includegraphics[height = 4.5cm]{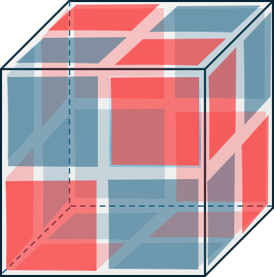}
 		\label{fig:s}}
                \subfloat[]{
 		\includegraphics[height = 4.5cm]{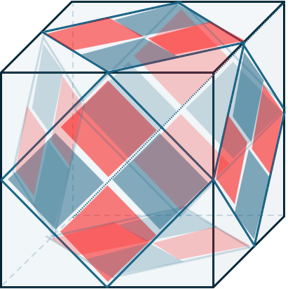}
 		\label{fig:s45}}
 		\caption{The two proposed point symmetric configurations. Left: the axes of the quadrupoles are parallel to the test mass edges; right: the axes of the quadrupoles are rotated by 45$^\circ$ with respect to the mass edges. Note the arrangement  of the magnets to maintain the full symmetry.}
\end{figure}

\begin{figure}[!h]
 		\centering
        \subfloat[]{
 		\includegraphics[height = 5cm]{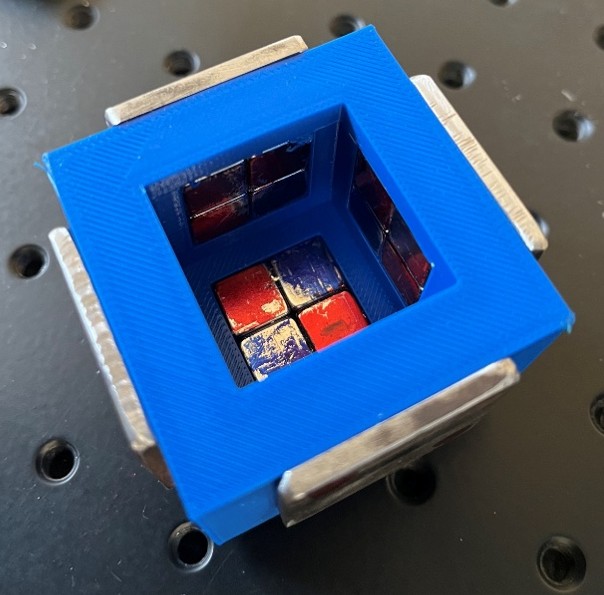}}
                \subfloat[]{
 		\includegraphics[height = 5cm]{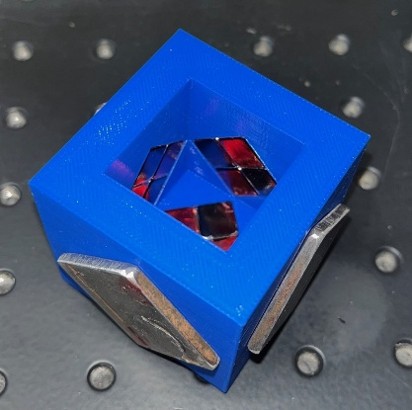}}
 		\caption{Experimental implementation of the  five-sided housing (left) and the 45$^\circ$  five-sided housing (right).}
        \label{fig:config}
\end{figure}

Two different configurations of the checkerboard magnetic pattern have been considered. In the first configuration (Figure \ref{fig:s}), the pattern is aligned with the cube's housing edges, whereas in the second (Figure \ref{fig:s45}), it is rotated by 45$^\circ$. The experimental results presented in the following sections are obtained using the first configuration; however, comparable results are also obtained with the 45$^\circ$  configuration, demonstrating that both configurations are effective in trapping the test mass.
Figure \ref{fig:config} shows the practical implementation of the magnetic trap realized with a plastic structure embedding five elementary quadruole units shown in Figure \ref{fig:fieldgradient}(a).

\begin{figure}[!h]
 		\centering
        \subfloat[]{
 		\includegraphics[height = 4cm]{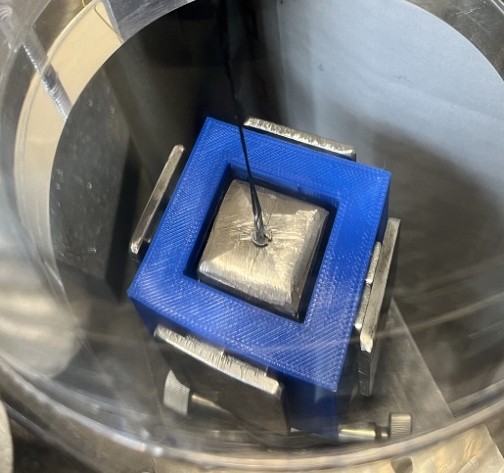}}
                \subfloat[]{
 		\includegraphics[height = 4cm]{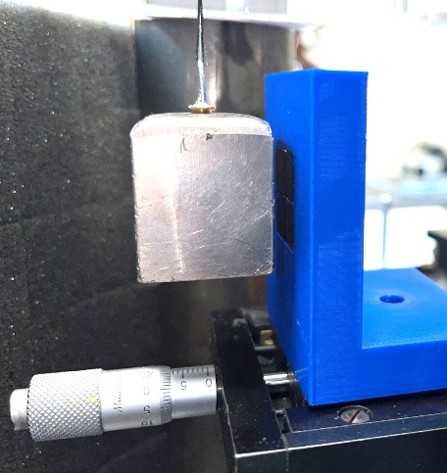}}
                \subfloat[]{
 		\includegraphics[height = 4cm]{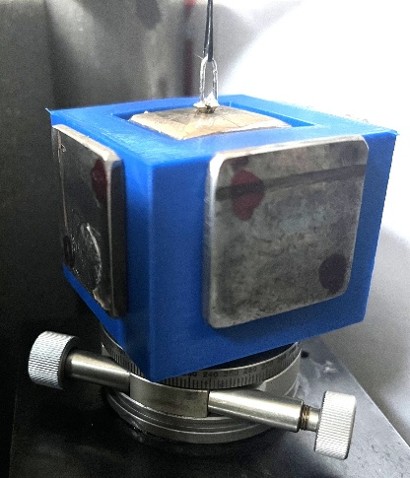}}
 		\caption{Pictures of the experimental setups. (a): suspended mass in the 5-sided housing seen from above where the gap is clearly visible. (b): experimental set-up for the measurement of the repulsion force where a single quadrupole is faced to the mass. (c): the housing placed on the rotation stage for the measurement of the torque.}
        \label{fig:exp}
\end{figure}

A ground-based experimental campaign is carried out using a simple pendulum testbed to validate and demonstrate the trapping concept. The suspension system consists of a 116 cm long 1K carbon-fiber wire supporting a solid bismuth parallelepiped with a square cross-section ($25\times25$ mm$^2$) and a height of approximately 30 mm, resulting in a test mass of about 180 g. The moment of inertia of the suspended mass about the rotation axis is $I_z = 1.91\times10^{-\mathbf{5}} \;\mathrm{kg\;m}^2$. %Given the experimentally observed rotational period of approximately 100 s, the torsional spring constant of the suspension wire is estimated to be $k = 9.20\times10^{-\mathbf{8}} \;\mathrm{N\;m/rad}$.
The pendulum motion is monitored using a Precitec optical displacement sensor (travel stroke equal of 10 mm and resolution 0.1 um), enabling accurate measurements of both translational displacements and angular oscillations. The pendulum configuration provides a 3 DoF system, in which the two horizontal translational degrees of freedom are counteracted by the gravitational restoring force, while rotation about the vertical axis is constrained by the torsional restoring torque of the suspension wire. The measured natural frequency of the pendulum modes is  0.47 Hz, while the torsional natural frequency is approximately  0.01 Hz, with both frequencies measured using the Precitec sensor.\\
Two key experiments are performed: the first to characterize the repulsive force and  estimate the resulting translational trapping force, and the second to characterize the magnetic torque responsible for maintaining the alignment of the test mass within the trap. Figure \ref{fig:exp} illustrates the different stages of the experimental procedures described below.

\subsection{Repulsion force and translational trapping}

\begin{figure}[!h]
 		\centering
 		\includegraphics[width = 0.6\columnwidth]{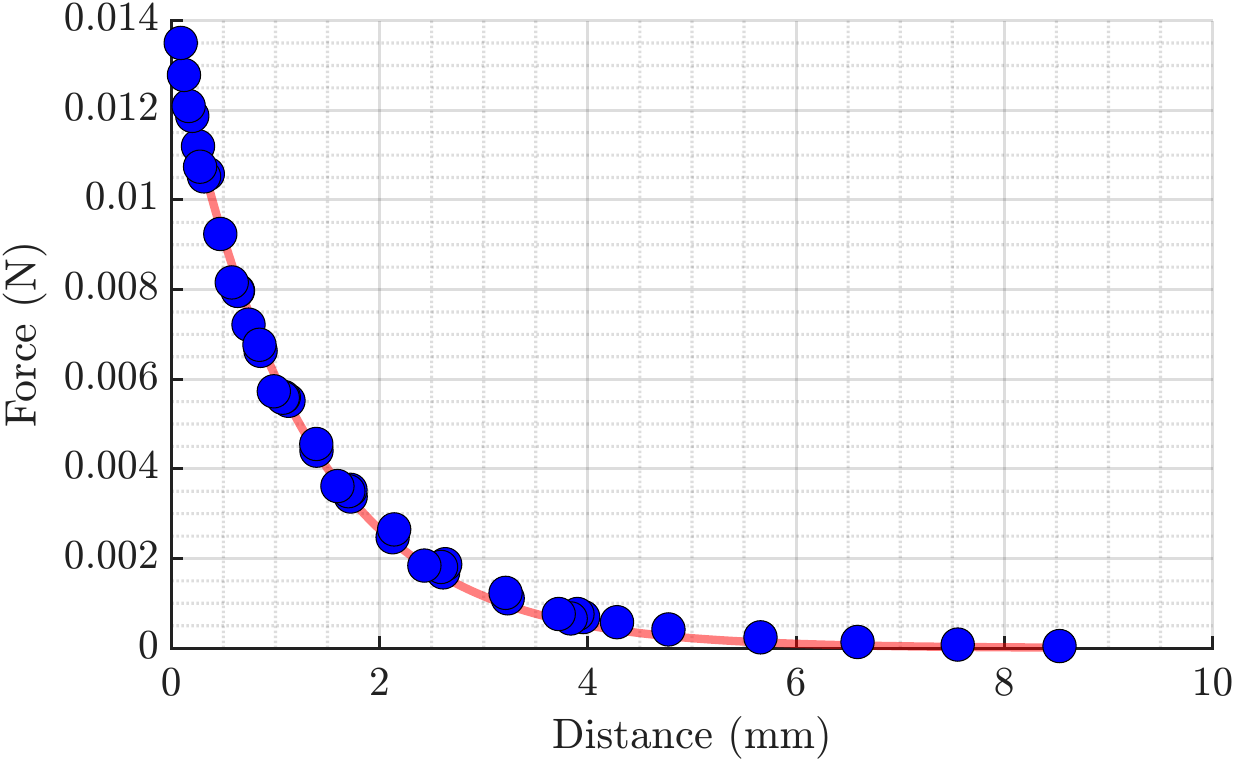}
 		\caption{Experimental results of the repulsive force as a function of distance. Blue dots: measured values; red curve: exponential fit.}
 		\label{fig:forcedist}
\end{figure}

To measure the repulsive force as a function of the gap between the quadrupole and the test mass, a single quadrupole is mounted vertically on a translation stage driven by a micrometric screw (Figure \ref{fig:exp} (b)). The quadrupole faces one of the vertical sides of the test mass whose displacement is measured by the Precitec sensor positioned on the opposite side of the mass (not visible in the picture). As the quadrupole is translated toward the mass, the repulsive interaction causes the mass to move away in the same direction. The gap is determined from the difference between the quadrupole displacement measured by the micrometric screw and the displacement of the mass measured by the Precitec sensor. The resulting displacement of the mass from its vertical equilibrium position is then used to determine the repulsive force from the horizontal component of the gravitational restoring force. The results of multiple measurements are reported in Figure \ref{fig:forcedist}. In the force/distance graph, the blue markers are the measured values, whereas the red curve is the exponential fit to the experimental data. The exponential fit obtained this way is used to estimate the translational trapping force produced by a pair of opposing quadrupoles.

\begin{figure}[!h]
 		\centering
        
 		\includegraphics[height = 5.7 cm]{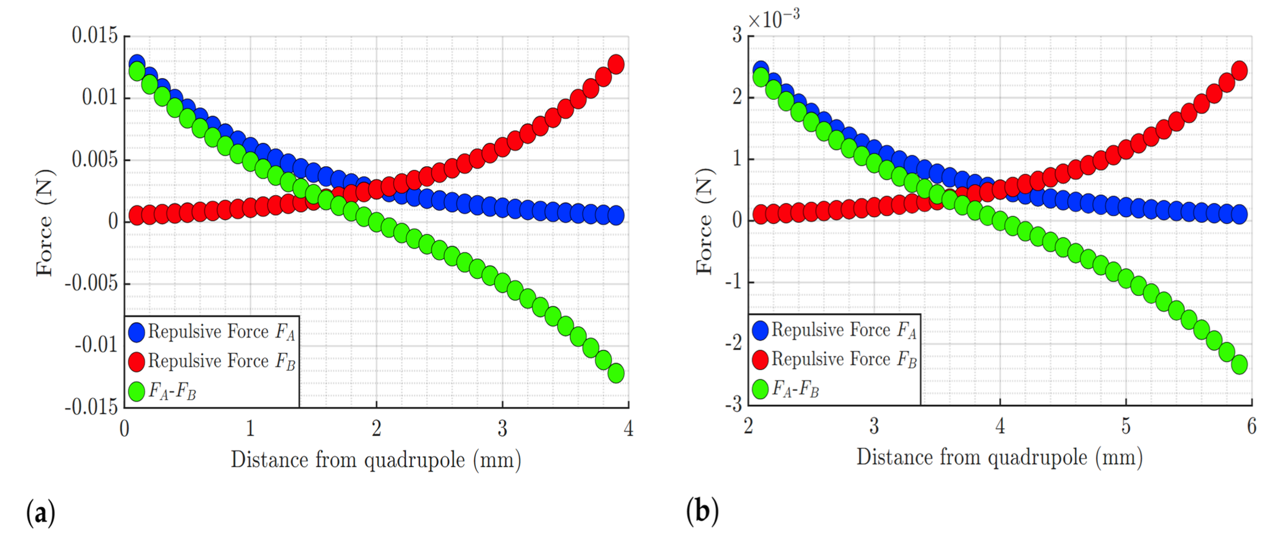}\\
 		\caption{Trapping force generated by a pair of opposite quadrupoles, denoted as $A$ and $B$, based on the experimentally determined force–gap relationship. The left and right plots correspond to gaps of 2 mm and 4 mm, respectively. The horizontal axis represents the distance between the test mass and quadrupole $A$. The blue and red curves show the repulsive forces generated by quadrupoles $A$ and $B$, respectively, while the green curve represents the resulting trapping force acting on the test mass, obtained as the difference between the two repulsive forces.}
        \label{fig:diff}
\end{figure}

 The trapping of the test mass is in fact generated by the repulsive forces from the quadrupoles on both sides, which balance each other at the central position. Figure \ref{fig:diff} shows the resulting net force acting on the test mass as a function of its position relative to the housing for the two considered configurations, corresponding to gaps of 2 mm and 4 mm. At the central position, the net force is zero, and the force–displacement relationship can be locally approximated as linear, corresponding to an ideal spring with an effective spring constant \textit{k} given by the derivative of the force with respect to displacement at the equilibrium position. The resulting spring constants are $k=4.38$ N$/$m and $k= 0.84$ N$/$m for the 2 mm and 4 mm gaps, respectively. In this estimation of the translational trapping force, the contributions from the other four quadrupoles, which move parallel to the faces of the cube, are neglected, as they are expected to be significantly smaller.

\subsection{Torque and rotational trapping}

Starting from the moment of inertia of the test mass and measuring the natural torsional period of the suspended mass (approximately 100 s), the torsional spring constant of the suspension wire can be estimated as $\kappa \approx 9.2 \times 10^{-8}$ N m$/$rad. As discussed below, this contribution is negligible compared with the magnetically induced restoring torque. The test mass is then inserted into the five-sided quadrupole housing, leaving the top open to accommodate the suspension wire. The housing is mounted on a rotation stage, as shown in Figure \ref{fig:exp} (c). When the housing is rotated, the test mass follows the rotation while maintaining its faces parallel to the housing walls, indicating the presence of a restoring magnetic torque. In order to quantify the value of the torque, a rapid angular displacement was manually applied to the housing, causing the test mass to oscillate about its axis without contacting the housing walls. The measured torsional natural frequencies are $f_{2mm} =
0.77$ Hz  and $f_{4mm} = 0.4$ Hz for gaps of 2 mm and 4 mm, respectively. These measurements allow the magnetically induced restoring torque to be estimated for the two configurations as $\kappa_{2mm}\approx 4.46\times {10}^{-4}$ N m$/$rad and $\kappa_{4mm} \approx 1.21\times {10}^{-4}$ N m$/$rad  for gaps of 2 mm and 4 mm, respectively. Both values are orders of magnitude larger than the torque of the wire, which, therefore, has been neglected. 

\subsection{Provisional sensitivity budget}

Based on the results obtained above, we can estimate a preliminary sensitivity budget for an accelerometer based on the proposed magnetic-trap principle. The acceleration sensitivity is determined by the resonant frequency of the mass–spring system and by the noise floor of the displacement readout. Assuming a displacement readout noise density of 100 pm$/$Hz, which can be readily achieved using either interferometric or capacitive sensing, the corresponding acceleration noise density can be estimated and combined with the thermal noise contribution. The resulting performance estimates for the proposed configuration are summarized in Table \ref{tab:noise_params} and provide a baseline for further developments.
The present analysis shows that the readout noise is the dominant contribution over the thermal (damping) noise. This is expected to change in high-end applications, where sub-picometer displacement readout can be employed, potentially enabling acceleration noise levels approaching $10^{-13} \text{ m/s}^2/\sqrt{\text{Hz}}$ \cite{LISA-INTERFEROMETER,pisani2018accelerometer}.\\
These preliminary results set the magnetically suspended accelerometer in a good position compared with presently available high-end accelerometers \cite{IAFOLLA2010,touboul1999electrostatic}.

\begin{table}[htbp]
\centering
\caption{Parameter values and noise sources for gap = 2 mm and gap = 4 mm.}
\label{tab:noise_params}
\begin{tabular}{p{6cm}ccc}
\toprule
\textbf{Parameter or noise source} & \textbf{Gap = 2 mm} & \textbf{Gap = 4 mm} & \textbf{Unit} \\
\midrule
\\
Translational magnetic stiffness: $k_{\mathrm{mag}}$
    & $4.38$ & $0.84$ & N/m \\
Translational natural frequency 
    & $0.78$ & $0.34$ & Hz \\
Torsional magnetic stiffness: $\kappa_{\mathrm{mag}}$
    & $4.46 \times 10^{-4}$ & $1.21 \times 10^{-4}$ & N$\cdot$m/rad \\
Torsional natural frequency
    & $0.77$ & $0.4$ & Hz \\
Equivalent acceleration noise (readout noise $= 10^{-10}$ m/$\sqrt{\mathrm{Hz}}$)
    & $2.39 \times 10^{-9}$ & $4.58 \times 10^{-10}$ & m/s$^2$/$\sqrt{\mathrm{Hz}}$ \\
Magnetic damping noise
    & $3.07 \times 10^{-10}$ & $1.61 \times 10^{-10}$ & m/s$^2$/$\sqrt{\mathrm{Hz}}$ \\
    \midrule
\textbf{Total acceleration noise}
    & $ \mathbf{2.41 \times 10^{-9}}$ & $\mathbf{4.85 \times 10^{-10}}$ & \textbf{ m/s}$\mathbf{^2}$\textbf{/}$\mathbf{\sqrt{Hz}}$ \\
\bottomrule
\end{tabular}
\end{table}

\section{Conclusions}

The proposed 6-quadrupole architecture is shown to constitute an effective passive alternative to mechanical and electrostatic suspensions, providing a stable and tunable magnetic potential well for the confinement of diamagnetic test masses. Ground-based pendulum measurements performed along three degrees of freedom (two translational and one rotational) demonstrate that the bismuth parallelepiped is stably trapped within the horizontal plane and naturally self-aligns with the walls of the magnetic system. Given the full symmetry of the 6-quadrupole configuration, the stability observed across these three degrees of freedom is expected to extend to the complete six degrees of freedom of a fully enclosed six-sided housing under microgravity conditions. The experimental results are in good agreement with the theoretical predictions obtained from numerical simulations developed in parallel.

Future work will address the following developments: the fabrication of a prototype with well defined geometrical parameters and a microcrystalline internal structure, aimed at mitigating anisotropy-induced effects; the repetition of the measurements under vacuum conditions to assess the damping coefficient; the realization of a full six-quadrupole assembly to be characterized in microgravity, for instance at the Einstein Elevator facility \cite{lotz2017einstein}; the implementation of an interferometric readout system capable of resolving picometer-level displacements and enabling a quantitative assessment of damping-induced noise; and the production of sintered or sectorized bismuth test masses designed to minimize thermal-magnetic damping. Taken together, these optimizations will pave the way toward the development of a flight-ready prototype. Further developments foresee the miniaturization of the magnetic trap to be compliant with the nano-sat market as well as the test of other diamagnetic materials such as highly oriented pyrolytic graphite (HOPG). 
%%%%%%%%%%%%%%%%%%%%%%%%%%%%%%%%%%%%%%%%%%

\authorcontributions{Conceptualization, writing---original draft preparation Marco Pisani; methodology, software, validation, formal analysis, writing---review and editing all authors.}

\funding{This research received no external funding.}

\institutionalreview{Not applicable.}

\acknowledgments{The authors aknowledge the colleague Lorenzo Giorio for the simulation of the magnetic field represented in figure 1, and the colleague Marco Santiano for the realization of the plastic supports.}

\conflictsofinterest{The authors declare no conflicts of interest.} 

%%%%%%%%%%%%%%%%%%%%%%%%%%%%%%%%%%%%%%%%%%
%% Optional

%% Only for journal Encyclopedia
%\entrylink{The Link to this entry published on the encyclopedia platform.}

\reftitle{References}

% Please provide either the correct journal abbreviation (e.g. according to the “The list of Title Word Abbreviations” https://portal.issn.org/ltwa) or the full name of the journal. 
% Citations and references in the Supplementary Materials are permitted provided that they also appear in the reference list here. 

% \bibliography{your_external_BibTeX_file}

%=====================================
% References, variant B: internal bibliography
%=====================================

%\bibliographystyle{IEEEtran}
\bibliography{biblio}

% If authors have biography, please use the format below
%\section*{Short Biography of Authors}
%\bio
%{\raisebox{-0.35cm}{\includegraphics[width=3.5cm,height=5.3cm,clip,keepaspectratio]{Definitions/author1.pdf}}}
%{\textbf{Firstname Lastname} Biography of first author}
%
%\bio
%{\raisebox{-0.35cm}{\includegraphics[width=3.5cm,height=5.3cm,clip,keepaspectratio]{Definitions/author2.jpg}}}
%{\textbf{Firstname Lastname} Biography of second author}

% For the MDPI journals use author-date citation, please follow the formatting guidelines on http://www.mdpi.com/authors/references
% To cite two works by the same author: \citeauthor{ref-journal-1a} (\citeyear{ref-journal-1a}, \citeyear{ref-journal-1b}). This produces: Whittaker (1967, 1975)
% To cite two works by the same author with specific pages: \citeauthor{ref-journal-3a} (\citeyear{ref-journal-3a}, p. 328; \citeyear{ref-journal-3b}, p.475). This produces: Wong (1999, p. 328; 2000, p. 475)

%%%%%%%%%%%%%%%%%%%%%%%%%%%%%%%%%%%%%%%%%%
%% for journal Sci
%\reviewreports{\\
%Reviewer 1 comments and authors’ response\\
%Reviewer 2 comments and authors’ response\\
%Reviewer 3 comments and authors’ response
%}
%%%%%%%%%%%%%%%%%%%%%%%%%%%%%%%%%%%%%%%%%%
\PublishersNote{}
%\isPreprints{}{% This command is only used for ``preprints''.
%\end{adjustwidth}
\end{document}